\documentclass[12pt]{article}
\usepackage{fullpage}
\usepackage{amsmath, amssymb, amsbsy, amsfonts, amsthm}
\usepackage[mathscr]{eucal}
\usepackage{bm}
\usepackage{ifthen}

\newcounter{repcount}
\newcommand{\Repeat}[2]{
  \ifthenelse{#1>0}{
  #2
  \setcounter{repcount}{#1}
  \addtocounter{repcount}{-1}
  \Repeat{\value{repcount}}{#2}}{}}

\theoremstyle{plain}
\newtheorem{theorem}{Theorem}
\newtheorem{lemma}{Lemma}
\theoremstyle{definition}

\theoremstyle{remark}

\theoremstyle{plain}
\newcommand{\startproof}{\noindent\textbf{Proof.} }
\newcommand{\finishproof}{\hfill $\blacksquare$ \\}

\newcommand{\mynote}[1]{}

\newcommand{\cfier}{C}

\newcommand{\C}{\mathbb{C}}

\newcommand{\R}{\mathbb{R}}

\newcommand{\dif}{\mathrm{d}}

\newcommand{\half}{\frac{1}{2}}

\newcommand{\nt}[1]{\mathring{#1}} 

\newcommand{\double}[2]{\hspace{0.1em} #1 \hspace{#2} #1 \hspace{0.1em}}
\newcommand{\ident}{\double{1}{-0.35em}}

\newcommand{\SO}{\mathrm{SO}}
\newcommand{\SU}{\mathrm{SU}}

\newcommand{\Hil}{\mathcal{H}}
\newcommand{\Cyl}{\mathrm{Cyl}}

\newcommand{\scrA}{\mathcal{A}}

\newcommand{\scrE}{\mathscr{E}}

\newcommand{\cylmid}{\big|}

\newcommand{\van}{\mathscr{V}}
\newcommand{\gact}{\rhd}
\newcommand{\tridec}[1]{\overset{\scriptstyle #1}{\rhd}}
\newcommand{\cact}{\tridec{C}}
\newcommand{\link}{\ell}

\newcommand{\BI}{\beta}

\newcommand{\mypara}[1]{\subsubsection*{#1}}

\begin{document}

\title{Embedding loop quantum cosmology without piecewise linearity}
\author{Jonathan Engle\thanks{jonathan.engle@fau.edu}
 \\[1mm]
\normalsize \em Department of Physics, Florida Atlantic University, Boca Raton, Florida, USA}
\date{\today}
\maketitle\vspace{-7mm}

\begin{abstract}

An important goal is to understand better the relation between full loop quantum gravity (LQG)
and the simplified, reduced theory known as loop quantum cosmology (LQC),
\textit{directly at the quantum level}.  Such a firmer understanding would increase confidence in the reduced theory
as a tool for formulating predictions of the full theory, as well as permitting lessons from the reduced theory to guide further development in the full theory.
The present paper constructs an embedding of the usual state space of LQC into that of standard LQG, that is, LQG
based on  \textit{piecewise analytic paths}.
The embedding is well-defined even prior to solving the diffeomorphism constraint, at no point is a graph fixed, and at no point is the piecewise linear category used.
This motivates for the first time a definition of operators in LQC corresponding to holonomies along non-piecewise-linear paths, without changing the usual kinematics of LQC in any way.  The new embedding intertwines all operators corresponding to such holonomies, and all elements in its image satisfy an operator equation which classically implies homogeneity and isotropy.
 The construction is made possible by a recent result proven by Fleischhack.

\end{abstract}



\section{Introduction}

Loop quantum gravity (LQG) is a well-defined, background independent framework for quantum gravity which admits well-defined quantizations of dynamics.  Loop quantum cosmology (LQC) is a simplified model which aims to describe
the cosmological consequences of LQG by attempting to model the homogeneous and isotropic sector of the theory.
 Concretely, this model is obtained by applying loop quantization methods to the classical symmetry reduced sector of gravity.
The simplicity of the symmetry reduced sector enables one to
complete the quantization procedure, and
begin the process of making predictions \cite{aps2006a, aps2006b, aps2006, gbcm2010, ahm2010, bct2011, ap2011, aan2012, aan2012a}.
In the course of developing LQC, lessons were learned, in particular
the need to change the quantization of the Hamiltonian constraint in order to recover the correct classical limit \cite{aps2006},
and it is hoped that these lessons will guide further development in full LQG.
Thus the value of LQC is two-fold:  It provides a way to derive predictions from LQG for cosmology, as well as providing a guide for further development in the full theory.
However, each of these goals can be achieved with full confidence and precision only insofar as LQC can be
\textit{derived} from LQG as the homogeneous isotropic sector in some sense.
The most obvious way of making precise the idea of LQC describing such a sector is to construct
an \textit{embedding} of LQC states into such a sector.
The meaning of the homogeneous-isotropic sector in a quantum field theory, and especially quantum gravity, was investigated in
the work \cite{engle2006}.  In the works \cite{engle2006, engle2007, engle2008},
a program, to a large extent successful, has been developed to construct an embedding into such a sector of LQG.
In doing this, at least until the present paper, it has been necessary ---
though only as an intermediate step \cite{engle2008} --- to first embed LQC into a version of LQG
in which only \textit{piecewise linear} paths are allowed \cite{engle2007, engle2008}.
The formal reason for this necessity
lies in a difference of choice made in the quantizations of LQG and LQC:
the choice of \textit{configuration algebra}, the algebra of functions on configuration space which are directly promoted to operators,
and which, when interpreted as states in the connection representation, additionally play the role of `test states' in the theory.
In standard LQG \cite{al2004, rovelli2004, thiemann2007}, this algebra includes
matrix elements of all parallel transports along \textit{all piecewise analytic paths}, whereas in LQC, this algebra includes
only matrix elements of parallel transports along paths which are \textit{adapted} to the homogeneous symmetry
group, in the sense that (piecewise) they are integrals of the corresponding Killing vector fields --- that is,
in the case of isotropic $k=0$ LQC, one restricts to
paths which are `piecewise-straight'.

In a recent paper \cite{fleischhack2010}, Fleischhack has begun to lay the foundations for an alternative construction of LQC
in which one uses the configuration algebra generated by parallel transports along all piecewise analytic paths,
as in the standard full theory.
This enables the embedding into standard LQG to be
achieved without using the piecewise linear category.
However, as straight forward as this sounds, it remains to see how far this alternative framework can be systematically
developed.
As of yet, no canonical inner product\footnote{The
work \cite{bk2010} mentions a possible inner product which, however, depends on a choice of
(Euclidean-group-equivalence class of ) three arbitrary fixed edges in space, among other choices.
}
or quantization of constraints and elementary observables of interest have been proposed.
The architects of the currently more well-established LQC kinematics
\cite{abl2003, aps2006} (which, for brevity, we refer to as `standard LQC') were in fact aware of the possibility of using a larger configuration algebra, but
chose to restrict consideration to the algebra generated by
linear paths in order to simplify the quantum theory and make it tractable
with the intuition and hope that this should be sufficient \cite{ashtekarprivate}.
%
%
For, one of the principal motivations for looking at loop quantum cosmology is precisely
to improve tractability of the tasks of completing, and computing
consequences from, the theory.

The present paper shows how, in fact,
\textit{standard} LQC \textit{can} be naturally embedded into standard, piecewise analytic LQG, without using piecewise linearity at any stage: \textit{a new formulation of LQC is not necessary} for this.
This embedding has been made possible by a technical result proven for the first time by Fleischhack in \cite{fleischhack2010}, a technical result which has also been addressed in the work \cite{bk2010}.
Specifically, the configuration algebra (and hence test states) proposed by Fleischhack \cite{fleischhack2010} separates cleanly into a direct sum of the configuration algebra of standard LQC and the space of functions vanishing at infinity and zero.

Of course, the embedding which we will introduce here is at the kinematical level.  In the end, as argued in \cite{engle2006, engle2007, engle2008},
it will be necessary to have an embedding at the diffeomorphism-invariant level, either by
making the idea sketched in \cite{engle2006, engle2007, engle2008} precise, or otherwise.

An added benefit of the new embedding of standard LQC is that it motivates a definition of
operators corresponding to holonomies along \textit{curved} paths
in \textit{standard} LQC, such that the new embedding 
intertwines these new operators with the corresponding operators in the full theory.
This was heretofore not possible.
Furthermore, closely related to this is that states in the image of the embedding satisfy
homogeneity and isotropy in a precise sense that does not use piecewise linearity in any way.
Thus, the embedding presented here has all the strengths of
that introduced in \cite{engle2007, engle2008}, and more.

Furthermore, the new embedding comes in two versions: a `c' embedding and a `b', or `holomorphic', embedding.  
As in \cite{engle2007}, the `c' embedding is simpler and introduced first, and the `b' embedding is defined in terms of it.
The characteristic difference between `c' and `b' embeddings lie in the way symmetry is satisfied by
states in their image. Physically, the role of an embedding is to identify the LQC state space with a particular `cosmological sector' of the full theory.  In the cosmological sector corresponding to `c' embeddings, only the configuration field satisfies homogeneity and isotropy, whereas in that of the `b' embeddings, \textit{both configuration and momenta} satisfy homogeneity and isotropy in
a precise sense. 
The use of coherent states, here and in \cite{engle2006, engle2007} 
specified by a choice of  positive phase space function called a `complexifier' \cite{thiemann2002}, is key in defining the `b' embeddings.
The cosmological sector of the `b' embedding presented in this paper,  when cut-off to a fixed graph and with a
particular choice of complexifier, is furthermore the same cosmological sector used in the spin-foam cosmology
literature \cite{rv2008,rv2009,brv2010,vidotto2010, vidotto2011}.
 However, in contrast to
the spin-foam cosmology literature, we fix no graph in this paper, and every state of the cosmological
sector considered here has support on every piecewise analytic graph, with no graph preferred.

The paper is organized as follows. We begin by reviewing the structures involved in LQG and LQC.  Fleischhack's
proposal and corresponding embedding are presented, and the intertwining and symmetry properties of this embedding
are proven.  This then provides a starting point for defining an embedding of \textit{standard} LQC into the full theory.
New operators in standard LQC for holonomies along curved
paths are motivated, and the intertwining property of these new operators using the new embedding
is proven.  The corresponding `b' or `holomorphic' embedding is then introduced, and all of the above properties
are proven also for it.  Lastly, in the discussion section,
the equivalence of the strategy of embedding pursued here and the strategy of specifying a projection
to relate LQC and LQG are briefly touched upon.

\section{LQG and LQC structures}

\mypara{Loop quantum gravity}
Loop quantum gravity is based on the \textit{Ashtekar-Barbero} formulation of gravity, in which the (unconstrained)
gravitational phase space $\Gamma$ is parameterized by
an $SU(2)$ connection $A_a \equiv A^i_a \tau_i$ and a densitized triad $\tilde{E}^a_i$ on space,
where $\tau_i := -\frac{i}{2}\sigma_i$ with $\sigma_i$ the Pauli matrices.
We use the convention that lower case latin indices are spatial indices.
The densitized triad field is related to a triad $e^a_i$ and its inverse $e^i_a$
via $\tilde{E}^a_i = \det (e^j_b) e^a_i$, which are related to the physical spatial metric via $q_{ab} = e_a^i e_{bi}$.
In terms of the generalized ADM variables, $A^i_a:=\Gamma^i_a + \BI K^i_a$,
where $\Gamma^i_a$ is the spin connection determined by $\tilde{E}^a_i$,
$K^i_a := K_{ab} e^{bi}$ with $K_{ab}$ the extrinsic curvature, and
$\BI \in \R^+$ is the Barbero-Immirzi parameter \cite{barbero1995, immirzi1997, meissner2004, abbdv2009, enpp2011, abk2000}.
%
%
%
%

The basic variables with direct quantum analogues are
holonomies $A(\link)$ of $A$ along piecewise analytic paths $\link$, and
electric fluxes through surfaces $S$:
$\Sigma(S)^i = \int_S \Sigma^i$,
where $\Sigma^i_{ab} := 2 \epsilon_{abc} \tilde{E}^{ci}$.
In the connection representation, states are wave functionals $\Psi(A)$ of the connection $A$. One starts with a space,
denoted $\Cyl$, of `nice' functions called \textit{cylindrical}, which
depend only the holonomies of $A$ along a finite set of (piecewise analytic) paths $\ell$;
when these paths are chosen to be non-intersecting except possibly at end points, they are called
\textit{edges} and their union is called a \textit{graph}. On $\Cyl$ is defined the
Ashtekar-Lewandowski inner product $\langle \cdot, \cdot \rangle$ \cite{ai1992, al1994};
the elementary operators $\widehat{A}(\link), \widehat{\Sigma}(S)^i$ then act on the resulting
completed  kinematical Hilbert space $\Hil_{kin}$.
The algebraic dual $\Cyl^*$ of $\Cyl$ provides a notion of distributional (i.e., non-normalizable) states,
elements of which we write with rounded bra notation \cite{afw2002}, $(\Psi \mid$.

\mypara{Loop quantum cosmology}

To define the homogeneous-isotropic sector (for the spatially flat case, which is the case we consider),
one fixes a specific action of the Euclidean
group, $\scrE \cong \R^3 \rtimes \SO(3)$, on the $\SU(2)$ principal fiber bundle of the
theory. Concretely, this is done by choosing an action $\gact$ of
$\scrE$
on the basic variables through spatial
diffeomorphisms and local $\SU(2)$ rotations:
\begin{equation}
\label{euclaction}
(x,r)\gact(A^i_a, \tilde{E}^a_i) = \left(r^i{}_j
\left(\phi_{(x,r)}\right)_* A^j_a \,\, , \,\, (r^{-1})^j{}_i
\left(\phi_{(x,r)}\right)_* \tilde{E}^a_j \right)
\end{equation}
where $r^i{}_j$ denotes the adjoint action of $r \in \SO(3)$.
If we let $\mathcal{A}_S$ and $\Gamma_S$ denote the set of elements of $\scrA$ and
$\Gamma$, respectively, fixed by this action, then $\scrA_S$ and $\Gamma_S$ are respectively
 one and two dimensional.  Fix a reference element $\nt{A}^i_a$ in $\scrA_S$ and
 a triad $\nt{e}^a_i$ such that
$\nt{e}^a_i \nt{A}^j_a = V_o^{-1/3} \delta^j_i$
for some $V_o$ with dimensions of volume. Let $\nt{q}_{ab}:=\nt{e}^i_a \nt{e}_{bi}$. Then
$(A^i_a, \tilde{E}^a_i) \in \Gamma_S$ if and only if it is of the form
\begin{displaymath}
A^i_a = c \nt{A}^i_a, \; \tilde{E}^a_i = p V_o^{-2/3} \sqrt{\nt{q}}
\nt{e}^a_i .
\end{displaymath}
For some real $c$ and $p$.
In the phase space $\Gamma_S$,
$c$ and $p$ are conjugate, and in the quantum theory one considers wavefunctions $\psi(c)$ of $c$.
A function $\psi(c)$ is called \textit{almost periodic} if it is a linear combination of
exponentials $e^{i\mu c}$
\cite{abl2003};
the space of such functions is denoted $\Cyl_S$.
In analogy with LQG one constructs an inner product $\langle , \rangle$ on $\Cyl_S$
and completes to obtain a Hilbert space $\Hil_S$.
Elements of $\Cyl_S^*$ represent distributional states and are again denoted using rounded bras
$(\psi |$.

\section{The Fleischhack state space}

We here review the configuration algebra, and hence `test states', proposed by Fleischhack \cite{fleischhack2010}, Brunnemann and Koslowski \cite{bk2010} for cosmology, as well as the embedding of these states
into full LQG mentioned in \cite{fleischhack2010}.  We also prove that this embedding satisfies all of the
properties satisfied by the embedding in \cite{engle2007}, except with the piecewise linear category
replaced by the piecewise analytic category. The proofs of these properties constitute
new results, though the arguments are strongly modeled on those already present in \cite{engle2007}.

\mypara{Definition and embedding}

The reference connection $\nt{A}_a$ provides a map $r: \R \rightarrow \scrA_S \subset \scrA$ via
\begin{equation}
\label{rdef}
r: c \mapsto c \mathring{A}_a.
\end{equation}
Let $\Cyl_F := r^*[\Cyl]$. This is  the configuration algebra, and hence the space of `test' states, which
Fleischhack and others \cite{fleischhack2010, bk2010} have advocated as an alternative foundation for quantum cosmology.
In such an alternative framework, $\Cyl^*_F$ would play the role of ``distributional states.'' The advantage of
such a framework lies in the existence of an embedding into full theory states,
$\iota_F : \Cyl_F^* \hookrightarrow \Cyl^*$, defined by
$(\iota_F \alpha | \Phi \rangle := (\alpha | r^* \Phi \rangle$. As we will see, $\iota_F$ is injective, thus justifying
the term `embedding.'  In fact, it is formally identical to the embedding defined in \cite{engle2007, engle2008},
with piecewise linearity replaced by piecewise analyticity.  As was
the case in \cite{engle2007, engle2008}, $\iota_F$ intertwines operators central to the quantizations, and its image satisfies an operator
equation implying homogeneity and isotropy, as we shall also prove.
\begin{lemma}
$\iota_F$ is injective and hence an embedding.
\end{lemma}
{\startproof

It is sufficient to show $\iota_F$ has trivial kernel. Suppose
$\iota_F \alpha = 0$. Then for all $\Phi \in \Cyl$,
\begin{displaymath}
0=(\iota_F  \alpha \cylmid  \Phi
\rangle = (\alpha \cylmid r^* \Phi \rangle.
\end{displaymath}
Because $r^*[\Cyl]=\Cyl_F$, the above implies
%
%
$\alpha = 0$.
\finishproof}

\mypara{Intertwining of operators}

Let $F(A)$ denote any cylindrical function, considered as a full theory phase space function depending only on $A$.
The restriction of $F(A)$ to the homogeneous isotropic sector is $F(r(c))$. Thus, in the quantum theory,
the full theory operator $\widehat{F(A)}$ corresponds to the reduced theory operator
$\widehat{F(r(c))}$.
We therefore adopt the notation
$\widehat{F(A)}_F := \widehat{F(r(c))}$. Using logic identical to that in proposition 2
of \cite{engle2007}
(which we do not repeat),
one proves the following.
\begin{theorem}
\label{finter}
$\iota_F$ intertwines
$\widehat{F(A)}_F$ and $\widehat{F(A)}$ in the
sense $\widehat{F(A)}^* \circ \iota_F = \iota_F \circ \widehat{F(A)}_F^*$.
\end{theorem}

\mypara{Homogeneity and isotropy}

States in the image of $\iota_F$ are furthermore in a precise sense homogeneous and isotropic.
To discuss homogeneity and isotropy at the quantum level, one must formulate
it in terms of holonomies and fluxes.
The condition of homogeneity and isotropy
on the connection takes the form \cite{engle2006, engle2007}
\begin{displaymath}
(g \gact A)(\ell)^A{}_B = A(\ell)^A{}_B,
\end{displaymath}
for all piecewise analytic $\ell$, all $g \in \mathscr{E}$, and all $A,B = 0,1$.

\begin{theorem}
\label{chom}
Every element $\Psi$ in the image of $\iota_F$ satisfies, for all piecewise analytic $\ell$ and $g \in \mathscr{E}$,
\begin{equation}
\label{symmop}
\widehat{(g \gact A)(\ell)^A{}_B}^* \Psi = \widehat{A(\ell)^A{}_B }^* \Psi .
\end{equation}
\end{theorem}
{\startproof

Let $g \in \mathscr{E}$, $\ell$, and $A,B \in \{0,1\}$ be given, and
let $F(A):= A(\ell)^A{}_B$ and $F'(A) := (g^* F)(A) = (g \gact A)(\ell)^A{}_B$.
As $r(c)$ is invariant under $\mathscr{E}$,
$F'(r(c)) := F(g \gact r(c)) = F(r(c))$.  Applying theorem \ref{finter}
to $\widehat{F(A)}$ and $\widehat{F'(A)}$ and equating yields
\begin{displaymath}
\widehat{F(A)}^* \iota_F \psi = \widehat{F'(A)}^* \iota_F \psi
\end{displaymath}
for all $\psi \in \Cyl_F^*$.
\finishproof}
Note this holds for all piecewise \textit{analytic} $\ell$, in contrast
to the piecewise linear result in \cite{engle2007}.

The fact that states in the image of $\iota_F$ satisfy a condition of
homogeneity and isotropy only on the connection
variable is due to the fact that it is a `c'-embedding (see \cite{engle2006, engle2007}).
Better in this respect are the `b' or holomorphic embeddings, in which symmetry is imposed on both configuration
and momenta by using coherent states \cite{engle2006, engle2007}. 
See section \ref{bsect}.

\section{Embedding of standard LQC into standard LQG}

We now come to the embedding of \textit{standard} LQC states into standard, piecewise analytic LQG
--- the central subject of this paper ---
and prove its properties.
This embedding is constructed using the embedding discussed above together with a recent decomposition of
$\Cyl_F$ proven by Fleischhack \cite{fleischhack2010}. We begin by reviewing this decomposition.

\mypara{Decomposition of the Fleischhack state space}
\label{fsect}

Let $\van$ denote the set of all functions on $\R$ vanishing at $\pm \infty$ and at zero. 
One then has the following result proven in \cite{fleischhack2010}:
\begin{theorem}
\label{decompthm}
As vector spaces,
\begin{equation}
\label{cylfdecomp}
\Cyl_F = \Cyl_S \oplus \van .
\end{equation}
\end{theorem}
\noindent This furthermore implies that $\Cyl_F^*$, the algebraic dual of $\Cyl_F$, physically representing `distributional states' in the framework proposed by Fleischhack, is naturally
isomorphic to $\Cyl_S^* \oplus \van^*$.
Let $P_S: \Cyl_F \rightarrow \Cyl_S$ and $P_V: \Cyl_F \rightarrow \van$ denote canonical projection onto the two
components in equation
(\ref{cylfdecomp}). 
\begin{lemma}
\label{starlemma}
Define $f: \Cyl_S^* \oplus \van^* \rightarrow \Cyl_F^*$ by
\begin{displaymath}
(f(\psi_S, \psi_V)| \phi \rangle := (\psi_S| P_S \phi \rangle + (\psi_V | P_V \phi \rangle .
\end{displaymath}
$f$ is one-to-one and onto, yielding a natural isomorphism $\Cyl_F^* \cong \Cyl_S^* \oplus \van^*$.
%
%
\end{lemma}
{\startproof

\noindent\textit{One-to-one:}

Suppose $f(\psi_S, \psi_V) = 0$.  Then for all $\phi \in \Cyl_S \subset \Cyl_F$,
$0=(f(\psi_S, \psi_V) | \phi \rangle = (\psi_S | \phi \rangle$, so that $\psi_S = 0$,
and for all $\phi \in \van \subset \Cyl_F$,
$0=(f(\psi_S, \psi_V) | \phi \rangle = (\psi_V | \phi \rangle$, so that $\psi_V = 0$.

\noindent\textit{Onto:}

Let $\psi \in \Cyl_F^*$ be given.  Define $\psi_S \in \Cyl_S^*$ and $\psi_V \in \van^*$
as the restriction of $\psi$ to $\Cyl_S \subset \Cyl_F$ and $\van \subset \Cyl_F$,
respectively.  Then $f(\psi_S,\psi_V) = \psi$.

%
%
%

\finishproof}

\mypara{Definition of the embedding}

Define $\iota_S: \Cyl_S^* \hookrightarrow \Cyl^*_F \cong \Cyl_S^* \oplus \van^*$
as the inclusion map via the isomorphism proved in lemma \ref{starlemma}.
%
%
One then defines $\iota: \Cyl_S^* \hookrightarrow \Cyl^*$ by
$\iota := \iota_F \circ \iota_S$.
Explicitly, for all $\psi \in \Cyl_S^*$ and $\Phi \in \Cyl$,
\begin{equation}
\label{iotaex}
(\iota \psi | \Phi \rangle = (\psi | P_S r^* \Phi \rangle .
\end{equation}
This is the embedding of central interest to this paper.
Because ${\rm Im} \iota \subset {\rm Im} \iota_F$,
\textit{elements in the image of $\iota$ are also homogeneous and isotropic
in the sense of theorem \ref{chom}} above.
Furthermore, as we will see below, $\iota$ satisfies direct analogues of
all other above properties of $\iota_F$ as well.

\mypara{New operators: Curved holonomies in standard LQC}

The fact that $\Cyl_S$ can be identified as a subspace of $\Cyl_F$
offers a method to define operators in standard LQC corresponding to holonomies
along curved paths, operators which heretofore were simply not defined.

The program initiated by Fleischhack \cite{fleischhack2010} does not yet include
the specification of an inner product on $\Cyl_F$.  However, let us suppose an inner product is chosen.
Let us furthermore assume that the restriction of the inner product on $\Cyl_F$ to $\Cyl_S$ is
the same as the usual Bohr inner product on $\Cyl_S$, and that $\van$ and $\Cyl_S$
are mutually orthogonal in this inner product\footnote{The usual way \cite{al1994a, ai1992, abl2003}
of constructing an inner product would be to use the Gel'fand transform \cite{gv1964}
%
%
to identify $\Cyl_F$ with
continuous functions on its Gel'fand spectrum \cite{fleischhack2010}
$\overline{\mathbb{R}}_{Bohr} \sqcup \mathbb{R}\setminus \{0\}$,
and then use an $L^2$ inner product determined by a choice of measure on this spectrum.
If one chooses the
measure on the Gel'fand spectrum  by combining the Haar measure on
$\overline{\mathbb{R}}_{Bohr}$ and any other measure on $\mathbb{R}\setminus \{0\}$,
the resulting inner product will satisfy the above assumptions.
An interesting question is whether the condition of invariance under residual diffeomorphisms used
in \cite{ac2012} would lead to such a measure in this case.
}.
Given an operator $\hat{O}_F$ on $\Hil_F$, one can then define a corresponding
operator $\hat{O}_S$ on $\Hil_S$ simply by matrix elements
\begin{equation}
\label{matel}
\langle \psi_S | \hat{O}_S | \phi_S \rangle
= \langle \psi_S | \hat{O}_F | \phi_S \rangle .
\end{equation}
From this one deduces
\begin{equation}
\label{proj}
\hat{O}_S = P_S \circ \hat{O}_F .
\end{equation}
Let $F(A)$ be any cylindrical function, considered as a full theory phase space function depending only on $A$,
and let $\widehat{F(A)}$ denote the corresponding quantum operator.
As noted earlier, the corresponding operator on $\Hil_F$ is $\widehat{F(A)}_F := \widehat{F(r(c))}$, so
that, from equation (\ref{proj}), the corresponding operator on the standard LQC Hilbert space $\Hil_S$
is given by $\widehat{F(A)}_S \equiv \widehat{F(r(c))}_S := P_S \circ \widehat{F(r(c))}$.

Let us apply this operator definition
to the matrix elements of the $SU(2)$ holonomy along an arbitrary piecewise analytic path $\ell$,
$F(A)^A{}_B := A(\ell)^A{}_B$.  The corresponding operator in standard LQC is then
\begin{equation}
\label{Shol}
(\widehat{A(\ell)}_S)^A{}_B := P_S \circ \widehat{(r(c))(\ell)}^A{}_B.
\end{equation}
To write this operator more explicitly and prove that is continues to behave as an $SU(2)$ holonomy,
we use the following key lemma which will again be important later.
\begin{lemma}
\label{homlemma} $P_S:\Cyl_F \to \Cyl_S$ is a multiplicative
homomorphism.
\end{lemma}
{\startproof

Let $f,g \in \Cyl_F$ be given. Then
\begin{eqnarray*}
fg &=& (P_Sf + P_V f)(P_S g + P_V g) \\
&=& (P_S f)(P_S g) + (P_Vf)(P_Vg) + (P_S f)(P_V g) + (P_V
f)(P_S g)
\end{eqnarray*}
As the first term is almost periodic and the last three terms vanish at infinity,
it follows
\begin{displaymath}
P_S(fg) = (P_S f)(P_S g) .
\end{displaymath}
\finishproof}
\textit{Remark:}
The key in the proof above is that any element of $\van$ times any element of $\Cyl_F$ is in
$\van$, that is, that $\van$ is an ideal.

With this lemma, the explicit action of the operator (\ref{Shol}) on an element $\psi \in \Cyl_S$ is
\begin{equation}
\label{Sholmult}
(\widehat{A(\ell)}_S)^A{}_B \psi(c) := P_S \left((r(c))(\ell)^A{}_B \psi(c)\right)
= \left(P_S (r(c))(\ell)^A{}_B \right) \psi(c)
\end{equation}
where $P_S \psi = \psi$ has been used.  Using again lemma \ref{homlemma} 
and the fact that $\overline{P_S f} = P_S \overline{f}$, it is straight forward to prove that
\begin{eqnarray*}
&& \left(\widehat{A(\ell)}_S\right)^A{}_B \left(\widehat{A(\ell)}_S^\dagger\right)^B{}_C =
\delta^A_C \ident, \\
&& \epsilon^{AB} \epsilon_{CD} \left(\widehat{A(\ell)}_S\right)^C{}_A  \left(\widehat{A(\ell)}_S\right)^D{}_B
= 2 \ident, \\
\text{and} \quad &&
 \left(\widehat{A(\ell \circ \ell')}_S\right)^A{}_C =
 \left(\widehat{A(\ell)}_S\right)^A{}_B  \left(\widehat{A(\ell')}_S\right)^B{}_C,
\end{eqnarray*}
for all piecewise analytic $\ell$ and $\ell'$, and
where, in the first equation, $\dagger$ denotes hermitian conjugation both as an operator and as a 2 by 2 matrix,
that is,
\begin{displaymath}
\left(\widehat{A(\ell)}_S^\dagger\right)^A{}_B := \left[\left(\widehat{A(\ell)}_S\right)^B{}_A\right]^\dagger .
\end{displaymath}
It follows that $\left(\widehat{A(\ell)}_S\right)^A{}_B$ indeed has eigenvalues only in $SU(2)$, and obeys
the composition law for parallel transports, as one would hope.

For completeness, we also give the explicit expression for $\left(\widehat{A(\ell)}_S\right)^A{}_B$ by writing out
explicitly the multiplicative factor in (\ref{Sholmult}).
Let a piecewise analytic path $\ell(t)$, where $t$ is an arc length parameter with respect to the background metric
$\mathring{q}$, be given. From equations (2-7) in \cite{fleischhack2010} and proposition 5.13 in \cite{fleischhack2010}, one deduces that the multiplicative factor is given
by\footnote{If
we identify $\nt{q}_{ab}$ in this paper with the background Euclidean metric in \cite{fleischhack2010, bf2007},
then $A_*$ in \cite{fleischhack2010, bf2007} becomes identified with $2 \nt{V}^{1/3} \nt{A}_a$ here, leading
to the extra factor of $2$ in the exponentials (\ref{multexp}) and the extra factor of $\nt{V}^{1/3}$ in the interpretation
of $\mu$ as compared with \cite{fleischhack2010}.}
\begin{equation}
\label{multexp}
P_S (r(c))(\ell) = A_+ e^{i\frac{\mu c}{2}} - A_- e^{-i\frac{\mu c}{2}}
\end{equation}
where $\nt{V}^{1/3}\mu$ is the geometric length of $\ell$ with respect to the background metric $\mathring{q}_{ab}$,
and the $A_\pm$ are two 2 by 2 matrices, independent of $c$, given
by
\begin{displaymath}
A_\pm = \frac{m(\nt{V}^{1/3}\mu)^{\half}}{2} e^{\pm \half R}
\left( \begin{array}{cc} m(0)^{-\half}(n(0) \pm 1) & m(0)^\half \\
- \overline{m(0)}^\half & \overline{m(0)}^{-\half} (n(0)\pm 1) \end{array}\right)
\end{displaymath}
where
\begin{eqnarray*}
\nt{V}^{1/3}\nt{A}_a \dot{\ell}^a &=:& -\frac{i}{2} \left( \begin{array}{cc} n & m \\ \overline{m} &  - n\end{array}\right),\\
R &:=& n(\nt{V}^{1/3}\mu) - n(0) - \int_0^{\nt{V}^{1/3}\mu} \frac{\dot{m}}{m} n \dif t,
\end{eqnarray*}
and the square root $m(t)^{1/2}$ is chosen such that it is continuous in $t$.
Note in particular that the multiplicative factor (\ref{multexp})  is \textit{not only almost periodic, but sinusoidal}
with period $4\pi/\mu$, just as in the piecewise straight case.

\mypara{Intertwining}

With the above definition of $\widehat{F(A)}_S$,
we show that the embedding $\iota: \Cyl_S^* \hookrightarrow \Cyl^*$
in (\ref{iotaex}) \textit{intertwines} the operators
$\widehat{F(A)}^*$ and $\widehat{F(A)}_S^*$ in full LQG and standard LQC,
giving yet further support
for \textit{both} the definition (\ref{matel}) of the operators on $\Hil_S$ as well as
the embedding $\iota$. 
\begin{theorem}
\label{iotainter}
$\iota$ intertwines
$\widehat{F(A)}_S^*$ and $\widehat{F(A)}^*$.
\end{theorem}
{\startproof
For all $\alpha \in \Cyl_S^*$ and $\Phi \in \Cyl$,
\begin{eqnarray*}
(\widehat{F(A)}^*  \iota \alpha \mid \Phi \rangle
&:=& (\iota \alpha \mid \widehat{F(A)} \Phi \rangle
:= (\alpha \mid P_S r^*(F(A) \Phi) \rangle
=  (\alpha \mid P_S(F(r(c)) \Phi(r(c))) \rangle \\
&=&  (\alpha \mid (P_S F(r(c))) P_S \Phi(r(c)) \rangle
=   (\alpha \mid \widehat{F(A)}_S P_S r^* \Phi \rangle
=   ( \widehat{F(A)}_S^* \alpha \mid P_S r^* \Phi \rangle \\
&=&  (\iota \widehat{F(A)}_S^* \alpha \mid \Phi \rangle
\end{eqnarray*}
where lemma \ref{homlemma} was used in the second line.
\finishproof}
This extends the intertwining result of \cite{engle2007} fully to the piecewise analytic
category without modifying the standard LQC Hilbert space $\Hil_S$ in anyway.

\mypara{Explicit expression in a common case}

In the case where the argument of the embedding $\iota$ is the dual  $(\theta^*|$ of an
element $\theta \in \Cyl_S$, a more direct expression is possible.
For all $f \in \Cyl_F$, define the \textit{mean}
\begin{displaymath}
M(f) := \lim_{T \rightarrow \infty} \frac{1}{2T} \int_{-T}^T f(c) \dif c .
\end{displaymath}
When $f$ is almost periodic, this is the
mean used by Bohr in \cite{bohr1947}.  The fact that it is well-defined also for $f \in \Cyl_F$
follows from the decomposition
\begin{equation}
\label{decomp}
f = P_S f + P_V f
\end{equation}
ensured by theorem  \ref{decompthm}:
because $P_V f$ vanishes at infinity, $M(f) = M(P_S f) + M(P_V f) =  M(P_S f)$.
For $\theta, \psi \in \Cyl_S$, $M$ is related to the inner product in standard LQC via
\begin{displaymath}
\langle \theta, \psi \rangle = M(\overline{\theta} \psi).
\end{displaymath}
Expression (\ref{iotaex}) then takes the form
\begin{eqnarray*}
\nonumber
\left( \iota   \theta^* | \Phi
\right\rangle &:=& ( \theta^* | P_S r^* \Phi \rangle
:= \langle \theta | P_S r^* \Phi \rangle
= M(\overline{\theta} P_S r^* \Phi)
= M(P_S(\overline{\theta} r^* \Phi)) \\
&=& M(\overline{\theta} r^* \Phi)
=  \lim_{T \rightarrow \infty} \frac{1}{2T} \int_{-T}^T \overline{\theta(c)} \Phi(r(c)) \dif c .
\end{eqnarray*}
In this expression, note that it is theorem \ref{decompthm}
which, by ensuring the decomposition (\ref{decomp}), ensures convergence of the limit.

\section{b-embeddings}
\label{bsect}

The embedding discussed thus far is of the type named `c'  in the
work \cite{engle2006, engle2007}, because the states in its image satisfy homogeneity
and isotropy only of the \textit{configuration field}.  To overcome this, and to
provide a better capacity to adapt to dynamics, the `b' embeddings were introduced \cite{engle2006, engle2007},
where `b' refers to the `balanced' way in which homogeneity and isotropy is imposed on both
configuration and momenta.

The basic idea  of the `b'-embeddings is to use \textit{coherent states} to
define an embedding of the reduced theory into the full theory.  In the work \cite{engle2007},
\textit{complexifier} coherent states \cite{thiemann2002} were used for this purpose.
These are directly related to a choice of complex coordinates on phase space.
The resulting embeddings,
 in contrast to the `c' embedding, intertwine an algebra of operators whose classical analogues \textit{separate points} on phase space, and the states in the image of each `b' embedding
satisfy an operator equation whose classical analogue implies homogeneity and isotropy of \textit{both} configuration and momentum fields. However, the `b' embeddings defined in \cite{engle2007} were limited to the piecewise linear category.  In the
present section, we apply ideas in \cite{engle2007} to the foregoing work of the present paper to obtain
a `b' embedding of LQC into \textit{piecewise analytic} LQG. Because most of the derivations
are formally the same as elsewhere, we skip almost all details and primarily state definitions and results.

\mypara{Definition}

The complex coordinates used in complexifier coherents states are generated
by a choice of positive function on phase space, called a \textit{complexifier}
\cite{thiemann2002}. 
Let $C: \Gamma \rightarrow \R^+$ and $C_S: \Gamma_S \rightarrow \R^+$
denote complexifiers on the full and reduced phase spaces of general relativity, and let
$\hat{C}$ and $\hat{C}_S$ denote their quantizations on $\Hil$ and $\Hil_S$,  respectively.
The corresponding classical complex coordinates
$\mathfrak{Z}$ and $z$ are then
\begin{equation}
\label{zetadef}
\mathfrak{Z}^i_a(x) := (\varphi_C(t)^* A^i_a(x))_{t \rightarrow i}
\qquad z := (\varphi_{C_S}(t)^* c)_{t \rightarrow i} .
\end{equation}
where $\varphi_C(t)$ and $\varphi_{C_S}(t)$
respectively denote the one parameter Hamiltonian flows on $\Gamma$ and $\Gamma_S$
generated by the phase space functions $C$ and $C_S$, and $t \rightarrow i$ denotes complex analytic continuation.
We make the same assumptions about $C$ and $C_S$ as were made in
\cite{engle2007}, namely (1.) they are pure momentum, $C=C[\tilde{E}^a_i]$, $C_S= C_S[p]$
(2.) $\frac{\delta C}{\delta \tilde{E}^a_i}$ and $\frac{d C_S}{d p}$ vanish only at $\tilde{E}^a_i = 0$
and $p =0$, and (3.) if $s: \C \rightarrow \mathcal{A}^\C$ denotes the inclusion map $\Gamma_S \hookrightarrow \Gamma$
in the coordinates $z$ and $\mathfrak{Z}$, then $s$ is holomorphic.
The necessary and sufficient conditions on $C$ and $C_S$ for $s$ to be holomorphic were derived in lemma 1
of the work \cite{engle2007}.
The above three conditions furthermore imply that $s$ is the analytic continuation of the map $r$ introduced in equation (\ref{rdef}).
%
%
%

We begin by stating the definition of the `b' embedding in terms of the `c' embedding.
Motivated by equation (68) 
%
%
of \cite{engle2007}, define $\iota_b: \Cyl_S^* \to \Cyl^*$ by
\begin{equation}
\label{bdef}
\iota_b := e^{-\hat{\cfier}^*}\circ \iota \circ
e^{\hat{\cfier}_S^*} .
\end{equation}
The injectivity of $\iota_b$ follows from the
injectivity of $\iota$.
$\iota_b$ furthermore maps complexifier coherent states to complexifier coherent states ---
see appendix \ref{bapp} for an exposition of this fact.
The other properties of $\iota_b$ are proven below, in turn.

\mypara{Intertwining of a set of operators whose classical analogues separate points}

Consider a function $F(\mathfrak{Z})$ depending holomorphically on a finite number of parallel transports of $\mathfrak{Z}$
along piecewise analytic paths --- i.e., a \textit{holomorphic, piecewise analytic cylindrical function}.
In contrast to the cylindrical functions of $A$, the  holomorphic cylindrical functions of $\mathfrak{Z}$ separate
points on $\Gamma$. From the fact that $F(\mathfrak{Z})$ is holomorphic, together with (\ref{zetadef}), one has
\begin{displaymath}
F(\mathfrak{Z}) = (\varphi_C(t)^* F(A))_{t \rightarrow i}
\end{displaymath}
from which follows the quantization\cite{thiemann2002}
%
%
\begin{displaymath}
\widehat{F(\mathfrak{Z})} = e^{\hat{C}} \widehat{F(A)} e^{-\hat{C}} .
\end{displaymath}
Because $s$ is holomorphic, $F(s(z))$ is likewise holomorphic, so that $F(s(z))$ can be quantized as an operator
on $\Hil_S$ in a similar way, yielding
\begin{equation}
\label{Fzop}
\widehat{F(s(z))}_S := e^{\hat{C}_S} \widehat{F(r(c))}_S e^{-\hat{C}_S}
 := e^{\hat{C}_S} \circ P_S \circ \widehat{F(r(c))} \circ e^{-\hat{C}_S}.
\end{equation}
\begin{theorem}
\label{binter} $\iota_b$ intertwines the dual action of any holomorphic, piecewise analytic cylindrical function $\widehat{F(\mathfrak{Z})}$ of $\mathfrak{Z}$, and the dual action of the corresponding reduced theory
operator $\widehat{F(s(z))}_S$:
\begin{displaymath}
\iota_b \circ \widehat{F(s(z))}_S^* = \widehat{F(\mathfrak{Z})}^* \circ \iota_b .
\end{displaymath}
\end{theorem}
{\startproof
The proof follows from theorem \ref{iotainter}, in the same way proposition 3
follows from proposition 2 in \cite{engle2007}. 
%
%
\finishproof}

\mypara{Homogeneity and isotropy}

Because $\mathfrak{Z}$ is a good coordinate on $\Gamma$, there exists a unique action $\cact$ of
the Euclidean group $\mathscr{E}$ on complex connections such that
\begin{equation}
g \cact \left(\mathfrak{Z}[p]\right) = \mathfrak{Z}[g \gact p]
\end{equation}
for all $g \in \mathscr{E}$ and $p \in \Gamma$. If the chosen complexifier $C$ is invariant under $\mathscr{E}$
(which will be the case if $C$ is diffeomorphism and $SU(2)$ gauge invariant), then
$\mathfrak{Z}[p]$ will transform covariantly under $\mathscr{E}$, and
the above action $\cact$ will be the same as  $\gact$  ---  i.e., for each $g \in \mathscr{E}$, 
$g \cact$ will act with the same combination of
diffeomorphisms and local $SU(2)$ rotations as in equation (\ref{euclaction}).

In terms of this action, the condition of homogeneity and isotropy
on the complex connection  $\mathfrak{Z}$ takes the form \cite{engle2006, engle2007}
\begin{displaymath}
(g \cact \mathfrak{Z})(\ell)^A{}_B = \mathfrak{Z}(\ell)^A{}_B ,
\end{displaymath}
for all piecewise analytic $\ell$, all $g \in \mathscr{E}$, and all $A,B = 0,1$.
This condition is equivalent to homogeneity of \textit{both the real connection $A^i_a$
and its conjugate momentum $\tilde{E}^a_i$}.  All states in the image of $\iota_b$ satisfy
the operator version of this condition:

\begin{theorem}
Every element $\Psi$ in the image of $\iota_b$ satisfies, for all piecewise analytic $\ell$ and $g \in \mathscr{E}$,
\begin{displaymath}
\widehat{(g \cact \mathfrak{Z})(\ell)^A{}_B}^* \Psi = \widehat{\mathfrak{Z}(\ell)^A{}_B }^* \Psi .
\end{displaymath}
\end{theorem}
{\startproof
For all $z$, $s(z)$ is by construction invariant under $\mathscr{E}$.
From this and equation (\ref{Fzop})
it follows that $\widehat{F'(s(z))}_S = \widehat{F(s(z))}_S$.
The proof then follows from
theorem \ref{binter} in the same way that theorem \ref{chom} followed from theorem \ref{finter} in section \ref{fsect}.
\finishproof}

\section{Discussion}

We have shown that it is not necessary at any point to restrict to the piecewise linear category
when embedding \textit{standard} loop quantum cosmology states into a homogeneous-isotropic sector
of full loop quantum gravity.  We have shown this by exhibiting such an embedding $\iota$,
not only of normalizable, but even of all \textit{distributional} states of standard LQC, $\Cyl_S^*$,
into standard LQG --- that is, LQG based on piecewise analytic graphs.
This has been done without fixing a graph.
The sense in which the image of the embedding consists in homogeneous
isotropic states is defined via operator equations.  Furthermore,
this embedding has motivated a new definition of operators in LQC for parallel transports along
\textit{curved} paths, heretofore undefined in LQC, which may be of use in applications.
These operators, together with the corresponding operators in the full theory,
are \textit{intertwined} by the embedding that has been introduced.
All of these results have been proven for both a `c' version and `b' --- or `holomorphic' --- version of the embedding,
in the terminology of \cite{engle2007}.

The properties of the embeddings introduced in this paper contrast with those of
the embedding suggested in \cite{bk2010}, whose image consists in cylindrical functions
based on a fixed graph with few edges
and which are far from homogeneous and isotropic.
Additionally, the embedding in \cite{bk2010}
has no property similar to the intertwining properties proven in theorems 4 and 6 of this paper.
However, states in the image of the embedding in \cite{bk2010} do consist in \textit{normalizable} states,
giving it the advantage of being usable in more contexts.

We have derived the embedding $\iota$ by starting from the embedding $\iota_F$ of an \textit{extended} space of states
$\Cyl_F^*$ into standard LQG.  This extended space of states has been proposed by
Fleischhack \cite{fleischhack2010} precisely because, by construction, it admits such an embedding.
Using a result of Fleischhack's \cite{fleischhack2010}, we have shown that $\Cyl_S^*$ is naturally isomorphic
to a subspace of $\Cyl_F^*$, so that by simply restricting $\iota_F$ to $\Cyl_S^*$, one obtains an embedding of 
$\Cyl_S^*$ into $\Cyl^*$,
which is precisely the $\iota$ we have introduced.
The injectivity of $\iota$ and the fact that states in its image are homogeneous and isotropic descend trivially
from the corresponding properties of $\iota_F$, which we have also proven here.
The fact that $\iota$ intertwines curved holonomies, on the other hand, is quite non-trivial and was by no means gauranteed.

One might take the viewpoint that $\Cyl_F$ is the more `fundamental' choice for the configuration algebra.
If one takes such a viewpoint, there are still heuristic arguments for why
the restriction to $\Cyl_S \subset \Cyl_F$ is appropriate and consistent.
The difference between $\Cyl_S$ and $\Cyl_F$
lay only in whether or not one includes holonomies along curved paths.
However, as was shown in \cite{engle2008},
in the full theory, exclusion of curved paths, once one solves the diffeomorphism constraint,
in fact does not alter the final theory\footnote{as
long as one uses a proposal for the extended diffeomorphism group which has already been advocated
on other grounds \cite{koslowski2006,  fr2004}.}. This suggests that, also in quantum cosmology, a restriction to
piecewise straight paths should be sufficient to capture all of the physics.
%
Such a restriction is furthermore consistent with the dynamics if quantized using the same strategy as that in the 
well-established `improved dynamics' quantization of \cite{aps2006}: It is easy to see that the resulting Hamiltonian
constraint operator $\hat{H}$ will preserve $\Cyl_S$ and hence its dual $\hat{H}^*$ will preserve 
%
%
$\Cyl_S^* \subset \Cyl_F^*$, 
so that one can consistently restrict to $\Cyl_S^*$.

We close with a note on the equivalence of the `embedding strategy' employed in this paper with the `projection strategy'
for relating LQC and LQG advocated in \cite{aw2009, as2011a}.
The present paper has considered the problem of relating a given quantum theory, with state space $\Hil$ and operators
$\hat{O}^i$, to some spatial symmetry reduction
thereof, with state space $\Hil_S$ and corresponding
operators $\hat{O}^i_S$.
In doing this, we have followed the general strategy of specifying an \textit{embedding} $\iota: \Hil_S \hookrightarrow \Hil$, such that the states in the image of the embedding satisfy operator equations expressing the relevant symmetry,
and hence membership in the corresponding `symmetric sector' of the full theory.
We wish to note that this strategy is fully equivalent
to the strategy suggested in the two papers
\cite{aw2009, as2011a}, in which one specifies a \textit{projection} from the larger space
of states $\Hil$ to the smaller $\Hil_S$, the interpretation being that of `integrating out the non-symmetric degrees of
freedom'.  This equivalence results from the fact that the adjoint of
every onto projection $\mathbb{P}: \Hil \rightarrow \Hil_S$ is injective, and hence an embedding
$\iota := \mathbb{P}^\dagger: \Hil_S \hookrightarrow \Hil$, and vice versa.
Furthermore, a pair of operators $\hat{O}, \hat{O}_S$ is intertwined by $\mathbb{P}$ if and only if their adjoints are
intertwined by the corresponding embedding $\iota = \mathbb{P}^\dagger$.
Indeed, one can see from (\ref{iotaex}) that the embedding we have proposed is the adjoint of the projection
$\mathbb{P} = P_S \circ r^*$.

We have taken the embedding viewpoint because it permits a clear sense in which homogeneity and isotropy
play a role. However, it should be emphasized that
the work \cite{aw2009} has achieved something important.  At least for the case of reducing the quantum
Bianchi I model to isotropic LQC,
the authors have constructed a \textit{dynamical projection}, which intertwines the \textit{Hamiltonian constraints} of
the two models, whence
the corresponding embedding also intertwines the Hamiltonian constraints.
LQC thus passes a first test of its ability to represent the dynamics of a less symmetric quantum model.
However, the role of homogeneity and isotropy in the definition of the dynamical projector does not have
a clear generalization to the full theory.
By contrast, the role of homogeneity and isotropy in the full theory embedding $\iota$ defined here is clear.
A hope is that if one understands better the relation between the dynamical projector of \cite{aw2009} 
and the strategy of embedding carried out in this paper, 
this might lead to a way to extend the success of \cite{aw2009} to the full theory.

\section*{Acknowledgements}

The author thanks Abhay Ashtekar for encouraging him to finish this work and Edward Wilson-Ewing
for discussions and remarks on a prior draft.
This work was supported in part by the
NSF through grant  PHY-1237510 and by
the National Aeronautics and Space Administration
through the University of Central Florida's NASA-Florida Space Grant Consortium.

\appendix

\section{Relation of the `b' embedding to coherent states}
\label{bapp}

In this appendix, we show a precise sense in which the embedding $\iota_b$, 
defined in equation (\ref{bdef}), maps coherent states to coherent states.
The complexifiers $\hat{C}$ and $\hat{C}_S$ determine
families of coherent states $\Psi^C_\mathfrak{Z} \in \Cyl$, $\psi^{C_S}_z \in \Cyl_S$ via
\begin{displaymath}
(\Psi^C_\mathfrak{Z} | \Phi \rangle := (e^{-\hat{C}}\Phi(A'))_{A' \rightarrow \mathfrak{Z}}
\qquad (\psi^{C_S}_z | \phi \rangle := (e^{-\hat{C}_S}\phi(c'))_{c' \rightarrow z}
\end{displaymath}
for all $\Phi \in \Cyl$ and $\phi \in \Cyl_S$, where `$\rightarrow$' denotes complex analytic continuation.
These are quantum states in the full and reduced theory, respectively, which are `peaked' at the
classical phase space points labeled by the coordinates $\mathfrak{Z}$ and $z$ \cite{thiemann2002}.

Given a graph $\gamma$, let $P_\gamma$
denote orthogonal projection, using the Ashtekar-Lewandowski inner product,
from $\Cyl$ to the space $\Cyl_\gamma$ of cylindrical functions which
depend only on holonomies along edges in $\gamma$.
Define truncations of $\iota_b$ and $\Psi_\mathfrak{Z}^C$ to the graph $\gamma$ by
${}^\gamma \iota_b := P_\gamma^* \circ \iota_b$
and ${}^\gamma \Psi^C_{\mathfrak{Z}}:= P_\gamma^*  \Psi^C_{\mathfrak{Z}}$ ---
this coincides with taking the `cut-off' \cite{thiemann2002} or `shadow' \cite{als2002}, on $\gamma$, of the states in the image of $\iota^b$, and of $\Psi^C_{\mathfrak{Z}}$.
%
%
\begin{theorem}
For $\gamma$ piecewise linear,
\begin{displaymath}
{}^\gamma \iota_b \psi_z^{C_S} = {}^\gamma \Psi^C_{s(z)} .
\end{displaymath}
\end{theorem}
{\startproof
For all $\Phi \in \Cyl$,
\begin{eqnarray*}
({}^\gamma \iota_b \psi_z^{C_S} | \Phi \rangle
&=& (\iota_b \psi_z^{C_S} | P_\gamma \Phi \rangle
= (\psi_z^{C_S} | e^{\hat{C}_S} P_S r^* e^{-\hat{C}} P_\gamma \Phi \rangle .
\end{eqnarray*}
Because $\hat{C}$ is pure momentum and hence graph preserving,
$e^{-\hat{C}}P_\gamma \Phi \in \Cyl_\gamma$ once more, so that
$r^* e^{-\hat{C}}P_\gamma \Phi \in \Cyl_\gamma$ is already in $\Cyl_S$, making
the projector unnecessary, whence
\begin{eqnarray*}
({}^\gamma \iota_b \psi_z^{C_S} | \Phi \rangle
&=& (\psi_z^{C_S} | e^{\hat{C}_S} r^* e^{-\hat{C}} P_\gamma \Phi \rangle
:= (r^* e^{-\hat{C}} P_\gamma \Phi)(c')_{c' \rightarrow z}
=  (e^{-\hat{C}} P_\gamma \Phi)(r(c'))_{c' \rightarrow z} \\
&=&  (e^{-\hat{C}} P_\gamma \Phi)(A')_{A' \rightarrow s(z)}
= (\Psi_{s(z)}^{C} | P_\gamma \Phi \rangle
= ({}^\gamma \Psi_{s(z)}^C | \Phi \rangle .
\end{eqnarray*}

\finishproof}

%

\begin{thebibliography}{10}

\bibitem{aps2006a}
A.~Ashtekar, T.~Pawlowski, and P.~Singh, ``{Quantum nature of the big bang},''
  {\em Phys. Rev. Lett.}, vol.~96, p.~141301, 2006.

\bibitem{aps2006b}
A.~Ashtekar, T.~Pawlowski, and P.~Singh, ``{Quantum nature of the big bang: An
  analytical and numerical investigation},'' {\em Phys. Rev.}, vol.~D73,
  p.~124038, 2006.

\bibitem{aps2006}
A.~Ashtekar, T.~Pawlowski, and P.~Singh, ``Quantum nature of the big bang:
  Improved dynamics,'' {\em Phys. Rev. D}, vol.~74, p.~084003, 2006.

\bibitem{gbcm2010}
J.~Grain, A.~Barrau, T.~Cailleteau, and J.~Mielczarek, ``Observing the big
  bounce with tensor modes in the cosmic microwave background: Phenomenology
  and fundamental {LQC} parameters,'' {\em Phys. Rev. D}, vol.~82, p.~123520,
  2010.

\bibitem{ahm2010}
C.~Afonso, A.~Henriques, and P.~Moniz, ``Inflation in loop quantum cosmology:
  Dynamics and spectrum of gravitational waves,'' {\em Phys. Rev. D}, vol.~81,
  p.~104049, 2010.

\bibitem{bct2011}
M.~Bojowald, G.~Calcagni, and S.~Tsujikawa, ``{Observational constraints on
  loop quantum cosmology},'' {\em Phys.Rev.Lett.}, vol.~107, p.~211302, 2011.

\bibitem{ap2011}
I.~Agullo and L.~Parker, ``{Stimulated creation of quanta during inflation and
  the observable universe},'' {\em Gen.Rel.Grav.}, vol.~43, pp.~2541--2545,
  2011.

\bibitem{aan2012}
I.~Agullo, A.~Ashtekar, and W.~Nelson, ``{A Quantum Gravity Extension of the
  Inflationary Scenario},'' {\em Phys.Rev.Lett.}, vol.~109, p.~251301, 2012.

\bibitem{aan2012a}
I.~Agullo, A.~Ashtekar, and W.~Nelson, ``{An Extension of the Quantum Theory of
  Cosmological Perturbations to the Planck Era},'' {\em arXiv:1211.1354}, 2012.

\bibitem{engle2006}
J.~Engle, ``Quantum field theory and its symmetry reduction,'' {\em Class.
  Quant. Grav.}, vol.~23, pp.~2861--2894, 2006.

\bibitem{engle2007}
J.~Engle, ``Relating loop quantum cosmology to loop quantum gravity: Symmetric
  sectors and embeddings,'' {\em Class. Quant. Grav.}, vol.~24, pp.~5777--5802,
  2007.

\bibitem{engle2008}
J.~Engle, ``Piecewise linear loop quantum gravity,'' {\em Class. Quant. Grav.},
  vol.~27, p.~035003, 2010.

\bibitem{al2004}
A.~Ashtekar and J.~Lewandowski, ``Background independent quantum gravity: A
  status report,'' {\em Class. Quant. Grav.}, vol.~21, 2004.

\bibitem{rovelli2004}
C.~Rovelli, {\em Quantum Gravity}.
\newblock Cambridge: Cambridge UP, 2004.

\bibitem{thiemann2007}
T.~Thiemann, {\em Modern Canonical Quantum General Relativity}.
\newblock Cambridge: Cambridge UP, 2007.

\bibitem{fleischhack2010}
C.~Fleischhack, ``Loop quantization and symmetry: Configuration spaces,'' {\em
  arXiv:1010.0449}, 2010.

\bibitem{bk2010}
J.~Brunnemann and T.~A. Koslowski, ``{Symmetry Reduction of Loop Quantum
  Gravity},'' {\em Class.Quant.Grav.}, vol.~28, p.~245014, 2011.

\bibitem{abl2003}
A.~Ashtekar, M.~Bojowald, and J.~Lewandowski, ``Mathematical structure of loop
  quantum cosmology,'' {\em Adv. Theor. Math. Phys.}, vol.~7, pp.~233--268,
  2003.

\bibitem{ashtekarprivate}
A.~Ashtekar, Private communication, 2005.

\bibitem{thiemann2002}
T.~Thiemann, ``Complexifier coherent states for quantum general relativity,''
  {\em Class. Quantum Grav.}, vol.~23, pp.~2063--2118, 2006.

\bibitem{rv2008}
C.~Rovelli and F.~Vidotto, ``Stepping out of homogeneity in loop quantum
  cosmology,'' {\em Class. Quant. Grav.}, vol.~25, p.~225024, 2008.

\bibitem{rv2009}
C.~Rovelli and F.~Vidotto, ``{On the spinfoam expansion in cosmology},'' {\em
  Class.Quant.Grav.}, vol.~27, p.~145005, 2010.

\bibitem{brv2010}
E.~Bianchi, C.~Rovelli, and F.~Vidotto, ``Towards spinfoam cosmology,'' {\em
  Phys. Rev. D}, vol.~82, p.~084035, 2010.

\bibitem{vidotto2010}
F.~Vidotto, ``{Spinfoam cosmology: Quantum cosmology from the full theory},''
  {\em J.Phys.Conf.Ser.}, vol.~314, p.~012049, 2011.

\bibitem{vidotto2011}
F.~Vidotto, ``{Many-nodes/many-links spinfoam: The homogeneous and isotropic
  case},'' {\em Class.Quant.Grav.}, vol.~28, p.~245005, 2011.

\bibitem{barbero1995}
J.~F. Barbero~G, ``Real ashtekar variables for lorentzian signature
  space-times,'' {\em Phys. Rev. D}, vol.~51, pp.~5507--5510, 1995.

\bibitem{immirzi1997}
G.~Immirzi, ``Real and complex connections for canonical gravity,'' {\em Class.
  Quant. Grav.}, vol.~14, pp.~L177--L181, 1995.

\bibitem{meissner2004}
K.~Meissner, ``Black-hole entropy in loop quantum gravity,'' {\em Class.
  Quantum Grav.}, vol.~21, pp.~5245--5251, 2004.

\bibitem{abbdv2009}
I.~Agullo, J.~F. Barbero, E.~Borja, J.~Diaz-Polo, and E.~Villasenor, ``The
  combinatorics of the {SU(2)} black hole entropy in loop quantum gravity,''
  {\em Phys. Rev. D}, vol.~80, p.~084006, 2009.

\bibitem{enpp2011}
J.~Engle, K.~Noui, A.~Perez, and D.~Pranzetti, ``The {SU(2)} black hole entropy
  revisited,'' {\em J. High Energy Phys.}, vol.~2011, p.~016, 2011.

\bibitem{abk2000}
A.~Ashtekar, J.~C. Baez, and K.~Krasnov, ``Quantum geometry of isolated
  horizons and black hole entropy,'' {\em Adv. Theor. Math. Phys.}, vol.~4,
  pp.~1--94, 2000.

\bibitem{ai1992}
A.~Ashtekar and C.~Isham, ``Representations of the holonomy algebras of gravity
  and non-abelian gauge theories,'' {\em Class. Quant. Grav.}, vol.~9,
  pp.~1433--1468, 1992.

\bibitem{al1994}
A.~Ashtekar and J.~Lewandowski, ``Differential geometry on the space of
  connections via graphs and projective limits,'' {\em J. Geom. Phys.},
  vol.~17, pp.~191--230, 1995.

\bibitem{afw2002}
A.~Ashtekar, S.~Fairhurst, and J.~Willis, ``Quantum gravity, shadow states, and
  quantum mechanics,'' {\em Class. Quant. Grav.}, vol.~20, pp.~1031--1062,
  2003.

\bibitem{al1994a}
A.~Ashtekar and J.~Lewandowski, ``{Projective techniques and functional
  integration for gauge theories},'' {\em J.Math.Phys.}, vol.~36,
  pp.~2170--2191, 1995.

\bibitem{gv1964}
I.~M. Gel'fand and N.~Y. Vilenkin, {\em Generalized Functions, Vol. 4:
  Applications of Harmonic Analysis}.
\newblock New York: Academic Press, 1964.

\bibitem{ac2012}
A.~Ashtekar and M.~Campiglia, ``{On the Uniqueness of Kinematics of Loop
  Quantum Cosmology},'' {\em Class.Quant.Grav.}, vol.~29, p.~242001, 2012.

\bibitem{bf2007}
J.~Brunnemann and C.~Fleischhack, ``{Non-almost periodicity of parallel
  transports for homogeneous connections},'' {\em Math.Phys.Anal.Geom.},
  vol.~15, pp.~299--315, 2012.
\newblock (\textit{Preprint: arXiv:0709.1621}, ``On the configuration spaces of
  homogeneous loop quantum cosmology and loop quantum gravity'').

\bibitem{bohr1947}
H.~Bohr, {\em Almost Periodic Functions}.
\newblock New York: Chelsea Publishing Company, 1947.

\bibitem{koslowski2006}
T.~A. Koslowski, ``{Physical diffeomorphisms in loop quantum gravity},'' {\em
  arXiv:gr-qc/0610017}, 2006.

\bibitem{fr2004}
W.~Fairbairn and C.~Rovelli, ``{Separable Hilbert space in loop quantum
  gravity},'' {\em J.Math.Phys.}, vol.~45, pp.~2802--2814, 2004.

\bibitem{aw2009}
A.~Ashtekar and E.~Wilson-Ewing, ``Loop quantum cosmology of {B}ianchi {I}
  models,'' {\em Phys. Rev. D}, vol.~79, p.~083535, 2009.

\bibitem{as2011a}
A.~Ashtekar and P.~Singh, ``{Loop quantum cosmology: A status report},'' {\em
  Class.Quant.Grav.}, vol.~28, p.~213001, 2011.

\bibitem{als2002}
A.~Ashtekar, J.~Lewandowski, and H.~Sahlmann, ``{Polymer and Fock
  representations for a scalar field},'' {\em Class.Quant.Grav.}, vol.~20,
  pp.~L11--1, 2003.

\end{thebibliography}
%
%
%
%


\end{document}